\documentclass[12pt]{article}
\usepackage{amsmath,amssymb}
\usepackage{graphicx}
\newcommand{\eqdef}{\stackrel{\text{def}}{=}}

\newcommand{\n}{\nonumber\\}

\newcommand{\ignore}[1]{}
\newcommand{\Romannumeral}[1]{\uppercase\expandafter{\romannumeral#1}}

\allowdisplaybreaks[4]

\begin{document}

\baselineskip=20pt

\newcommand{\preprint}{
\vspace*{-20mm}
   \begin{flushright}\normalsize \sf
    DPSU-20-2\\
  \end{flushright}}
\newcommand{\Title}[1]{{\baselineskip=26pt
  \begin{center} \Large \bf #1 \\ \ \\ \end{center}}}
\newcommand{\Author}{\begin{center}
  \large \bf Satoru Odake \end{center}}
\newcommand{\Address}{\begin{center}
     Faculty of Science, Shinshu University,
     Matsumoto 390-8621, Japan
   \end{center}}
\newcommand{\Accepted}[1]{\begin{center}
  {\large \sf #1}\\ \vspace{1mm}{\small \sf Accepted for Publication}
  \end{center}}

\preprint
\thispagestyle{empty}

\Title{Free Oscillator Realization of the Laguerre Polynomial}

\Author

\Address
\vspace{5mm}

\begin{abstract}
We revisit the radial oscillator from the free oscillator realization point
of view. By using a free oscillator, namely the creation/annihilation operators
of the harmonic oscillator, we construct an operator that maps the
eigenfunctions of the harmonic oscillator to those of the radial oscillator.
As a polynomial part of this relation, we obtain an operator that maps the
Hermite polynomials to the Laguerre polynomials.
\end{abstract}

\section{Introduction}
\label{sec:intro}

Free field realization is a useful and powerful tool for physicists to study
complicated mathematical objects. For example, infinite dimensional algebras
such as the (deformed) Virasoro algebra, correlation functions of solvable
lattice models and eigenfunctions of exactly solvable many body quantum
mechanical systems etc.\ are studied by free field realization, see e.g.
\cite{o99}.
Roughly speaking, a free field is an infinite collection of free oscillators,
which are the creation/annihilation operators of the harmonic oscillator.
To study quantum mechanical systems with one degree of freedom by free field
approach, a free field is not necessary and it is sufficient to use a free
oscillator, namely, free oscillator realization.

In this letter, we study free oscillator realization of the radial oscillator.
Eigenfunctions of the harmonic oscillator are described by the Hermite
polynomial, and those of the radial oscillator are described by the Laguerre
polynomial.
By using a free oscillator, we construct some operator.
We show that the eigenfunctions of the radial oscillator are obtained from
those of the harmonic oscillator by this operator.
As a polynomial part of this relation, the Laguerre polynomial is obtained
from the Hermite polynomial.
We also show the orthogonality relation of the Laguerre polynomial by using
the properties of the Hermite polynomial.

This letter is organized as follows.
The harmonic oscillator and the Hermite polynomial are recapitulated in
\S\,\ref{sec:HO}, and the radial oscillator and the Laguerre polynomial are
recapitulated in \S\,\ref{sec:RO}
In \S\,\ref{sec:FOR} free oscillator realization of the radial oscillator is
considered. The eigenfunctions of the radial oscillator are obtained from
those of the harmonic oscillator.
Section \ref{sec:summary} is for a summary and comments.

\section{Harmonic Oscillator}
\label{sec:HO}

In this section we recapitulate the harmonic oscillator \cite{cks} and the
Hermite polynomial \cite{kls}.

The Hamiltonian of the harmonic oscillator is
\begin{equation}
  \mathcal{H}=p^2+x^2-1,
\end{equation}
(the standard normalization is
$\mathcal{H}^{\text{st}}=\frac12(\mathcal{H}+1)$), where $x\in\mathbb{R}$ is
the coordinate and $p=-i\frac{d}{dx}$ is the momentum.
The Schr\"odinger equation and eigenfunctions are
\begin{align}
  &\mathcal{H}\phi_n(x)=\mathcal{E}_n\phi_n(x)\ \ (n\in\mathbb{Z}_{\geq 0}),
  \label{Scheq}\\
  &\mathcal{E}_n=2n,\quad
  \phi_n(x)=e^{-\frac12x^2}H_n(x),\quad
  (\phi_m,\phi_n)=2^nn!\sqrt{\pi}\,\delta_{mn},
\end{align}
where the inner product of functions $f$ and $g$ is
$(f,g)\eqdef\int_{-\infty}^{\infty}\!dxf(x)^*g(x)$.
Here $H_n(x)$ is the Hermite polynomial,
\begin{equation}
  H_n(x)=(2x)^n{}_2F_0\Bigl(\genfrac{}{}{0pt}{}{-\frac{n}{2},-\frac{n-1}{2}}{-}
  \Bigm|-\frac{1}{x^2}\Bigr)
  =n!\sum_{k=0}^{[\frac{n}{2}]}\frac{(-1)^k(2x)^{n-2k}}{k!\,(n-2k)!},
  \label{Hn}
\end{equation}
where $[x]$ denotes the greatest integer not exceeding $x$ and ${}_rF_s$ is
the hypergeometric function,
\begin{equation}
  {}_rF_s\Bigl(\genfrac{}{}{0pt}{}{a_1,\ldots,a_r}{b_1,\ldots,b_s}\Bigm|z\Bigr)
  =\sum_{k=0}^{\infty}\frac{(a_1,\ldots,a_r)_k}{(b_1,\ldots,b_s)_k}
  \frac{z^k}{k!}.
\end{equation}
More explicitly, \eqref{Hn} is
\begin{alignat}{2}
  H_{2n}(x)&=\sum_{k=0}^nc'_{n,k}x^{2k},&\quad
  c'_{n,k}&=\frac{(-1)^{n-k}2^{2k}(2n)!}{(2k)!\,(n-k)!},
  \label{H2n}\\
  H_{2n+1}(x)&=\sum_{k=0}^nc''_{n,k}x^{2k+1},&\quad
  c''_{n,k}&=\frac{(-1)^{n-k}2^{2k+1}(2n+1)!}{(2k+1)!\,(n-k)!}.
  \label{H2n+1}
\end{alignat}
Note that $2^{-n}H_n(x)$ is a monic polynomial in $x$.

Annihilation and creation operators $a$ and $a^{\dagger}$, and the number
operator $N$ are defined by
\begin{equation}
  a\eqdef\frac{1}{\sqrt{2}}(x+ip),\quad
  a^{\dagger}\eqdef\frac{1}{\sqrt{2}}(x-ip),\quad
  N\eqdef a^{\dagger}a,
\end{equation}
and they satisfy
\begin{equation}
  [a,a^{\dagger}]=1,\quad[N,a]=-a,\quad[N,a^{\dagger}]=a^{\dagger},\quad
  \mathcal{H}=2N\ \ \bigl(\mathcal{H}^{\text{st}}=N+\tfrac12\bigr),
\end{equation}
and
\begin{equation}
  a\phi_n(x)=\sqrt{2}\,n\phi_{n-1}(x),\quad
  a^{\dagger}\phi_n(x)=\frac{1}{\sqrt{2}}\,\phi_{n+1}(x),\quad
  N\phi_n(x)=n\phi_n(x).
  \label{aphin}
\end{equation}
By the similarity transformation in terms of the ground state wavefunction
$\phi_0(x)=e^{-\frac12x^2}$, we define $\tilde{X}$ for an operator $X$ as
\begin{equation}
  \tilde{X}\eqdef\phi_0(x)^{-1}\circ X\circ\phi_0(x).
  \label{tildeX}
\end{equation}
For example, $\tilde{x}=x$, $\tilde{p}=p+ix$,
$\tilde{a}=\frac{1}{\sqrt{2}}\frac{d}{dx}$,
$\tilde{a^{\dagger}}=\frac{1}{\sqrt{2}}(-\frac{d}{dx}+2x)$ and
$\tilde{N}=\tilde{a^{\dagger}}\tilde{a}$.
The relations \eqref{aphin} become
\begin{equation}
  \tilde{a}H_n(x)=\sqrt{2}\,nH_{n-1}(x),\quad
  \tilde{a^{\dagger}}H_n(x)=\frac{1}{\sqrt{2}}\,H_{n+1}(x),\quad
  \tilde{N}H_n(x)=nH_n(x).
  \label{taphin}
\end{equation}

The first equation of \eqref{taphin} is
\begin{equation}
  \frac{d}{dx}H_n(x)=2nH_{n-1}(x).
  \label{dHn}
\end{equation}
{}From \eqref{H2n}--\eqref{H2n+1}, we have
\begin{equation}
  \frac{H_{2n+1}(x)}{2^{2n+1}x}=\sum_{k=0}^n\frac{(-1)^kn!}{(n-k)!}
  \frac{H_{2(n-k)}(x)}{2^{2(n-k)}}.
  \label{Hodd_byeven}
\end{equation}
By \eqref{dHn}, we have
$\frac{d}{dx}\phi_{2n}(x)=4n\phi_{2n-1}(x)-x\phi_{2n}(x)$.
Combining this and \eqref{Hodd_byeven}, we obtain
\begin{equation}
  \frac{1}{x}\frac{d}{dx}\frac{\phi_{2n}(x)}{2^{2n}}
  =-2\sum_{k=0}^n\frac{(-1)^kn!}{(n-k)!}
  \frac{\phi_{2(n-k)}(x)}{2^{2(n-k)}}+\frac{\phi_{2n}(x)}{2^{2n}}.
  \label{1/xdphi2n}
\end{equation}

For later use, let us define an inner product of functions $f$ and $g$ on
$\mathbb{R}_{>0}$,
\begin{equation}
  (f,g)'\eqdef\int_0^{\infty}\!dxf(x)^*g(x).
  \label{(,)'}
\end{equation}
We remark that the eigenfunctions $\phi_{2n}(x)$ and $\phi_{2n+1}(x)$ are also
orthogonal functions with respect to this inner product,
\begin{equation}
  (\phi_{2m},\phi_{2n})'=2^{2n-1}(2n)!\sqrt{\pi}\,\delta_{mn},\quad
  (\phi_{2m+1},\phi_{2n+1})'=2^{2n}(2n+1)!\sqrt{\pi}\,\delta_{mn}.
\end{equation}

\section{Radial Oscillator and Free Oscillator Realization}
\label{sec:ROFOR}

In this section, after recapitulating the radial oscillator \cite{cks} and
the Laguerre polynomial \cite{kls}, we present its free oscillator realization.

\subsection{Radial oscillator}
\label{sec:RO}

The Hamiltonian of the radial oscillator is
\begin{equation}
  \mathcal{H}^{\text{L}}=p^2+x^2+\frac{g(g-1)}{x^2}-2g-1,
  \label{HL}
\end{equation}
where the coordinate $x$ takes a value in $\mathbb{R}_{>0}$ and $g$ is a
coupling constant ($g>\frac12$).
The eigenfunctions of the Schr\"odinger equation \eqref{Scheq} (with superscript
$\text{L}$) are
\begin{equation}
  \mathcal{E}^{\text{L}}_n=4n,\quad
  \phi^{\text{L}}_n(x)=x^ge^{-\frac12x^2}L^{(g-\frac12)}_n(x^2),\quad
  (\phi^{\text{L}}_m,\phi^{\text{L}}_n)'
  =\frac{1}{2n!}\Gamma(n+g+\tfrac12)\delta_{mn}.
  \label{phiLn}
\end{equation}
Here $L^{(\alpha)}_n(\eta)$ is the Laguerre polynomial,
\begin{equation}
  L^{(\alpha)}_n(\eta)=\frac{(\alpha+1)_n}{n!}
  {}_1F_1\Bigl(\genfrac{}{}{0pt}{}{-n}{\alpha+1}\Bigm|\eta\Bigr)
  =\frac{1}{n!}\sum_{k=0}^n\frac{(-n)_k}{k!}(\alpha+k+1)_{n-k}\eta^k.
  \label{Ln}
\end{equation}
Note that $(-1)^nn!\,L^{(\alpha)}_n(\eta)$ is a monic polynomial in $\eta$.
By the similarity transformation in terms of the factor $x^g$,
$\mathcal{H}^{\text{L}}$ becomes
\begin{equation}
  (x^g)^{-1}\circ\mathcal{H}^{\text{L}}\circ x^g
  =-\frac{d^2}{dx^2}-\frac{2g}{x}\frac{d}{dx}+x^2-2g-1
  =\mathcal{H}-2g-\frac{2g}{x}\frac{d}{dx}.
  \label{tgHL}
\end{equation}

For later use, we present the following identities with a parameter $g$
($n,m\in\mathbb{Z}_{\geq 0}$):
\begin{align}
  &\sum_{k=0}^n\frac{(g)_k}{k!}=\frac{(g+1)_n}{n!},
  \label{id1}\\
  &\sum_{k=0}^{m-l}\genfrac{(}{)}{0pt}{}{m}{k}
  \frac{(2\bigl(m-k)\bigr)!}{2^{2(m-k)}(m-k-l)!}(g)_k
  =\frac{(2l)!}{2^{2l}}\genfrac{(}{)}{0pt}{}{m}{l}
  \frac{(g+\frac12)_m}{(g+\frac12)_l}\quad(0\leq l\leq m),
  \label{id2}\\
  &\sum_{l_1=0}^m\sum_{l_2=0}^n(-1)^{l_1+l_2}
  \genfrac{(}{)}{0pt}{}{m}{l_1}\genfrac{(}{)}{0pt}{}{n}{l_2}
  \frac{(g+\frac12)_{l_1+l_2}}{(g+\frac12)_{l_1}(g+\frac12)_{l_2}}
  =\delta_{mn}\frac{n!}{(g+\frac12)_n}.
  \label{id3}
\end{align}

\subsection{Free oscillator realization}
\label{sec:FOR}

Let us define an operator $b$ as
\begin{equation}
  b\eqdef\frac{1}{N+1}a^2.
\end{equation}
By \eqref{aphin}, action of $b$ on $\phi_n(x)$ is
\begin{equation}
  b\phi_0(x)=b\phi_1(x)=0,\quad
  b\phi_n(x)=2n\phi_{n-2}(x)\ \ (n\geq 2),
\end{equation}
and we have
\begin{equation}
  b^k\frac{\phi_{2n}(x)}{2^{2n}}
  =\frac{n!}{(n-k)!}\frac{\phi_{2(n-k)}(x)}{2^{2(n-k)}}
  \ \ (0\leq k\leq n),\quad
  b^k\phi_{2n}(x)=0\ \ (k>n).
  \label{bkphi2n}
\end{equation}

Let us consider the following function:
\begin{equation}
  F_n(x)\eqdef x^g\,{}_1F_0\Bigl(\genfrac{}{}{0pt}{}{g}{-}\Bigm|-b\Bigr)
  \frac{\phi_{2n}(x)}{2^{2n}}\ \ (n\in\mathbb{Z}_{\geq 0}).
  \label{Fndef}
\end{equation}
Although the hypergeometric function ${}_1F_0$ is an infinite series, it is
truncated to a finite sum by \eqref{bkphi2n},
\begin{equation}
  F_n(x)=x^g\sum_{k=0}^n\frac{(g)_k}{k!}(-b)^k\frac{\phi_{2n}(x)}{2^{2n}}
  =x^g\sum_{k=0}^n(-1)^k\genfrac{(}{)}{0pt}{}{n}{k}(g)_k\,
  \frac{\phi_{2(n-k)}(x)}{2^{2(n-k)}},
  \label{Fn}
\end{equation}
which is square integrable with respect to $(\,,\,)'$.
Action of $\mathcal{H}^{\text{L}}$ on $F_n(x)$ is
\begin{align}
  \mathcal{H}^{\text{L}}F_n(x)
  &=x^g\cdot x^{-g}\mathcal{H}^{\text{L}}x^g\cdot
  \sum_{k=0}^n(-1)^k\genfrac{(}{)}{0pt}{}{n}{k}(g)_k
  \frac{\phi_{2(n-k)}(x)}{2^{2(n-k)}}\n
  &\stackrel{(\text{\romannumeral1})}{=}
  x^g\sum_{k=0}^n(-1)^k\genfrac{(}{)}{0pt}{}{n}{k}(g)_k
  \Bigl(\mathcal{H}-2g-\frac{2g}{x}\frac{d}{dx}\Bigr)
  \frac{\phi_{2(n-k)}(x)}{2^{2(n-k)}}\n
  &\stackrel{(\text{\romannumeral2})}{=}
  x^g\sum_{k=0}^n(-1)^k\genfrac{(}{)}{0pt}{}{n}{k}(g)_k\,
  4(n-k-g)\frac{\phi_{2(n-k)}(x)}{2^{2(n-k)}}\n
  &\quad
  +x^g\sum_{k=0}^n(-1)^k\genfrac{(}{)}{0pt}{}{n}{k}(g)_k\,
  4g\sum_{l=0}^{n-k}\frac{(-1)^l(n-k)!}{(n-k-l)!}
  \frac{\phi_{2(n-k-l)}(x)}{2^{2(n-k-l)}},
  \label{HLFn}
\end{align}
where we have used (\romannumeral1): \eqref{tgHL},
(\romannumeral2): \eqref{Scheq} and \eqref{1/xdphi2n}.
By changing the summation variable $k,l$ to $k,m=k+l$,
the second term of \eqref{HLFn} (in the last line) becomes
\begin{align*}
  &\quad x^g\sum_{m=0}^n\sum_{k=0}^m(-1)^k\genfrac{(}{)}{0pt}{}{n}{k}(g)_k\,
  4g(-1)^{m-k}\frac{(n-k)!}{(n-m)!}
  \frac{\phi_{2(n-m)}(x)}{2^{2(n-m)}}\\
  &=x^g\sum_{m=0}^n(-1)^m\genfrac{(}{)}{0pt}{}{n}{m}
  \frac{\phi_{2(n-m)}(x)}{2^{2(n-m)}}
  4g\,m!\sum_{k=0}^m\frac{(g)_k}{k!}\\
  &\stackrel{(\text{\romannumeral1})}{=}
  x^g\sum_{m=0}^n(-1)^m\genfrac{(}{)}{0pt}{}{n}{m}
  \frac{\phi_{2(n-m)}(x)}{2^{2(n-m)}}4g(g+1)_m\\
  &=x^g\sum_{m=0}^n(-1)^m\genfrac{(}{)}{0pt}{}{n}{m}(g)_m\,
  4(g+m)\frac{\phi_{2(n-m)}(x)}{2^{2(n-m)}},
\end{align*}
where we have used \eqref{id1} in (\romannumeral1).
Therefore we obtain
\begin{equation}
  \mathcal{H}^{\text{L}}F_n(x)
  =x^g\sum_{k=0}^n(-1)^k\genfrac{(}{)}{0pt}{}{n}{k}(g)_k\,
  4n\frac{\phi_{2(n-k)}(x)}{2^{2(n-k)}}
  =4nF_n(x).
\end{equation}
This means that $F_n(x)$ is an eigenfunction of $\mathcal{H}^{\text{L}}$
with energy eigenvalue $4n$, namely, $F_n(x)\propto\phi^{\text{L}}_n(x)$.
By comparing the highest term of the polynomial part (the part other than
$x^ge^{-\frac12x^2}$), we obtain
\begin{equation}
  (-1)^nn!\,\phi^{\text{L}}_n(x)=F_n(x),
  \label{phiLn=Fn}
\end{equation}
namely,
\begin{equation}
  (-1)^nn!\,\phi^{\text{L}}_n(x)
  =x^g\,{}_1F_0\Bigl(\genfrac{}{}{0pt}{}{g}{-}\Bigm|-b\Bigr)
  \frac{\phi_{2n}(x)}{2^{2n}}.
  \label{phiLn=xgFphi2n}
\end{equation}
Therefore the eigenfunction of the radial oscillator $\phi^{\text{L}}_n(x)$
is obtained from that of the harmonic oscillator $\phi_{2n}(x)$ by the operator
$x^g\,{}_1F_0\bigl(\genfrac{}{}{0pt}{}{g}{-}\!\!\bigm|\!\!-b\bigr)$.
The polynomial part of this relation gives
\begin{equation}
  (-1)^nn!\,L^{(g-\frac12)}_n(x^2)
  ={}_1F_0\Bigl(\genfrac{}{}{0pt}{}{g}{-}\Bigm|-\tilde{b}\Bigr)
  \frac{H_{2n}(x)}{2^{2n}},
  \label{Ln=1F0H2n}
\end{equation}
where $\tilde{b}=\frac{1}{\tilde{N}+1}\tilde{a}^2$.
This gives the Laguerre polynomial in terms of the Hermite polynomial of even
degree,
\begin{equation}
  (-1)^nn!\,L^{(g-\frac12)}_n(x^2)
  =\sum_{k=0}^n(-1)^k\genfrac{(}{)}{0pt}{}{n}{k}(g)_k
  \frac{H_{2(n-k)}(x)}{2^{2(n-k)}}.
  \label{Ln=sumHeven} 
\end{equation}

Next let us consider the orthogonality relation of $F_n(x)$.
For $k\in\mathbb{Z}_{\geq 0}$, we have
\begin{equation}
  \int_0^{\infty}\!dxx^{2g}e^{-x^2}x^{2k}=\frac12\Gamma(g+k+\tfrac12)
  =(g+\tfrac12)_k\frac{\Gamma(g+\frac12)}{2}.
  \label{intx2k}
\end{equation}
The inner product $(F_m,F_n)'$ is calculated as follows:
\begin{align}
  &\quad(F_m,F_n)'\n
  &\stackrel{(\text{\romannumeral1})}{=}
  \sum_{k_1=0}^m(-1)^{k_1}\genfrac{(}{)}{0pt}{}{m}{k_1}(g)_{k_1}
  \sum_{k_2=0}^n(-1)^{k_2}\genfrac{(}{)}{0pt}{}{n}{k_2}(g)_{k_2}
  \Bigl(x^g\frac{\phi_{2(m-k_1)}(x)}{2^{2(m-k_1)}},
  x^g\frac{\phi_{2(n-k_2)}(x)}{2^{2(n-k_2)}}\Bigr)'\n
  &\stackrel{(\text{\romannumeral2})}{=}
  \sum_{k_1=0}^m(-1)^{k_1}\genfrac{(}{)}{0pt}{}{m}{k_1}(g)_{k_1}
  \sum_{k_2=0}^n(-1)^{k_2}\genfrac{(}{)}{0pt}{}{n}{k_2}(g)_{k_2}
  \sum_{l_1=0}^{m-k_1}\sum_{l_2=0}^{n-k_2}
  \frac{c'_{m-k_1,l_1}}{2^{2(m-k_1)}}\frac{c'_{n-k_2,l_2}}{2^{2(n-k_2)}}
  (g+\tfrac12)_{l_1+l_2}\frac{\Gamma(g+\frac12)}{2}\n
  &\stackrel{(\text{\romannumeral3})}{=}
  \sum_{l_1=0}^m\sum_{l_2=0}^n
  \frac{(-1)^{m+n-l_1-l_2}2^{2(l_1+l_2)}}{(2l_1)!\,(2l_2)!}
  (g+\tfrac12)_{l_1+l_2}\n
  &\quad\times
  \sum_{k_1=0}^{m-l_1}\genfrac{(}{)}{0pt}{}{m}{k_1}
  \frac{(2\bigl(m-k_1)\bigr)!}{2^{2(m-k_1)}(m-k_1-l_1)!}(g)_{k_1}
  \sum_{k_2=0}^{n-l_2}\genfrac{(}{)}{0pt}{}{n}{k_2}
  \frac{(2\bigl(n-k_2)\bigr)!}{2^{2(n-k_2)}(n-k_2-l_2)!}(g)_{k_2}
  \times\frac{\Gamma(g+\frac12)}{2}\n
  &\stackrel{(\text{\romannumeral4})}{=}
  \sum_{l_1=0}^m\sum_{l_2=0}^n(-1)^{l_1+l_2}
  \genfrac{(}{)}{0pt}{}{m}{l_1}\genfrac{(}{)}{0pt}{}{n}{l_2}
  \frac{(g+\frac12)_{l_1+l_2}}{(g+\frac12)_{l_1}(g+\frac12)_{l_2}}
  \times(-1)^{m+n}(g+\tfrac12)_m(g+\tfrac12)_n
  \frac{\Gamma(g+\frac12)}{2}\n
  &\stackrel{(\text{\romannumeral5})}{=}
  \delta_{mn}n!\,(g+\tfrac12)_n\frac{\Gamma(g+\frac12)}{2}
  =\frac12n!\,\Gamma(g+n+\tfrac12)\delta_{mn},
\end{align}
where we have used (\romannumeral1): \eqref{Fn},
(\romannumeral2): \eqref{H2n} and \eqref{intx2k},
(\romannumeral3): \eqref{H2n} and changing of the order of sums,
(\romannumeral4): \eqref{id2},
(\romannumeral5): \eqref{id3}.
On the other hand, \eqref{phiLn=Fn} gives
$(F_m,F_n)'=(-1)^{m+n}m!\,n!\,(\phi^{\text{L}}_m,\phi^{\text{L}}_n)'$.
Thus the orthogonality relation
$(\phi^{\text{L}}_m,\phi^{\text{L}}_n)'
=\frac{1}{2n!}\Gamma(n+g+\frac12)\delta_{mn}$ \eqref{phiLn} is recovered.

\section{Summary and Comments}
\label{sec:summary}

Free oscillator realization of the radial oscillator is studied.
We have shown that the operator
$x^g\,{}_1F_0\bigl(\genfrac{}{}{0pt}{}{g}{-}\!\!\bigm|\!\!-b\bigr)$ maps
the eigenfunction of the harmonic oscillator $\phi_{2n}(x)$ to that of the
radial oscillator $\phi^{\text{L}}_n(x)$, \eqref{phiLn=xgFphi2n}.
As the polynomial part of this relation, the operator
${}_1F_0\bigl(\genfrac{}{}{0pt}{}{g}{-}\!\!\bigm|\!\!-\tilde{b}\bigr)$ maps
the Hermite polynomial $H_{2n}(x)$ to the Laguerre polynomial
$L^{(g-\frac12)}_n(x^2)$, \eqref{Ln=1F0H2n}.
This gives the Laguerre polynomial in terms of the Hermite polynomial of even
degree, \eqref{Ln=sumHeven}.
The orthogonality relation of the Laguerre polynomial is obtained by using
the properties of the Hermite polynomial.

We have presented the operator that maps $\phi_{2n}(x)$ to
$\phi^{\text{L}}_n(x)$.
Similarly, we can construct an operator, which maps $\phi_{2n+1}(x)$ to
$\phi^{\text{L}}_n(x)$.
The result is as follows:
\begin{equation}
  (-1)^nn!\,\phi^{\text{L}}_n(x)
  =x^{g-1}\,{}_1F_0\Bigl(\genfrac{}{}{0pt}{}{g-1}{-}\Bigm|-b'\Bigr)
  \frac{\phi_{2n+1}(x)}{2^{2n+1}},\quad
  b'\eqdef\frac{1}{N+2}a^2.
  \label{phiLn=F'n}
\end{equation}
The polynomial part of this relation gives
\begin{equation}
  (-1)^nn!\,L^{(g-\frac12)}_n(x^2)
  =\frac{1}{x}\,{}_1F_0\Bigl(\genfrac{}{}{0pt}{}{g-1}{-}\Bigm|-\tilde{b'}\Bigr)
  \frac{H_{2n+1}(x)}{2^{2n+1}},
  \label{Ln=1F0H2n+1}
\end{equation}
where $\tilde{b'}=\frac{1}{\tilde{N}+2}\tilde{a}^2$.
This gives the Laguerre polynomial in terms of the Hermite polynomial of
odd degree,
\begin{equation}
  (-1)^nn!\,L^{(g-\frac12)}_n(x^2)
  =\sum_{k=0}^n(-1)^k\genfrac{(}{)}{0pt}{}{n}{k}(g-1)_k
  \frac{H_{2(n-k)+1}(x)}{2^{2(n-k)+1}x}.
  \label{Ln=sumHodd}
\end{equation}
Inverse transformations of \eqref{Ln=sumHeven} and \eqref{Ln=sumHodd} are
\begin{align}
  \frac{H_{2n}(x)}{2^{2n}}
  &=\sum_{k=0}^n\genfrac{(}{)}{0pt}{}{n}{k}(g+1-k)_k\,
  (-1)^{n-k}(n-k)!\,L^{(g-\frac12)}_{n-k}(x^2),
  \label{H2n=Ln}\\
  \frac{H_{2n+1}(x)}{2^{2n+1}x}
  &=\sum_{k=0}^n\genfrac{(}{)}{0pt}{}{n}{k}(g-k)_k\,
  (-1)^{n-k}(n-k)!\,L^{(g-\frac12)}_{n-k}(x^2)
  \label{H2n+1=Ln}\\
  &\stackrel{(\text{\romannumeral1})}{=}
  \sum_{k=0}^n\genfrac{(}{)}{0pt}{}{n}{k}(g+1-k)_k\,
  (-1)^{n-k}(n-k)!\,L^{(g+\frac12)}_{n-k}(x^2),
  \label{H2n+1=Ln2}
\end{align}
where we have shifted $g\to g+1$ in (\romannumeral1) because \eqref{H2n+1=Ln}
does not depend on $g$.
We remark that by setting $g=0$ in \eqref{H2n=Ln} and \eqref{H2n+1=Ln2},
they give the well known formula \cite{kls},
\begin{equation}
  H_{2n}(x)=(-1)^nn!\,2^{2n}L^{(-\frac12)}_n(x^2),\quad
  H_{2n+1}(x)=(-1)^nn!\,2^{2n+1}xL^{(\frac12)}_n(x^2).
\end{equation}

There are infinitely many exactly solvable deformations of the radial
oscillator, which are described by the multi-indexed Laguerre orthogonal
polynomials \cite{os25}.
It is an interesting problem to study their free oscillator realizations.

\section*{Acknowledgements}

This work was supported by JSPS KAKENHI Grant Number JP19K03667.


\end{document}